\documentclass[a4paper,fleqn]{cas-dc}

\usepackage[numbers,sort&compress]{natbib}
\usepackage{url}
\usepackage{hyperref}
\usepackage{amsmath}
\usepackage[final]{microtype}
\usepackage{placeins}

\ExplSyntaxOn
\cs_gset:Npn \__first_footerline: {}
\ExplSyntaxOff

\def\tsc#1{\csdef{#1}{\textsc{\lowercase{#1}}\xspace}}
\tsc{WGM}
\tsc{QE}
\tsc{EP}
\tsc{PMS}
\tsc{BEC}
\tsc{DE}

\begin{document}
\let\WriteBookmarks\relax
\def\floatpagepagefraction{1}
\def\textpagefraction{.001}
\shorttitle{fitPALSpectra}
\shortauthors{G. E. Pavlou}

\title [mode = title]{fitPALSpectra: Python fitting of positron annihilation lifetime spectra}

\author[1,2]{Georgios E. Pavlou}[orcid=0000-0003-4478-935X]
\ead{gepavlou@iesl.forth.gr}
\ead[url]{https://github.com/gepavlou}


\affiliation[1]{organization={Institute of Electronic Structure and Laser (IESL), FORTH},
                postcode={71110}, 
                city={Heraklion},
                country={Greece}}

\affiliation[2]{organization={Department of Physics, University of Crete},
                postcode={70013}, 
                city={Heraklion},
                country={Greece}}

\begin{abstract}
Positron annihilation lifetime spectroscopy (PALS) spectra are commonly analyzed by fitting multi-exponential lifetime models convoluted with the detector resolution function. In practice, this inverse problem is sensitive to initial parameter choices, parameter bounds, source corrections, and correlations between lifetime and intensity parameters. This paper presents fitPALSpectra, an open-source Python workflow for configurable PALS spectrum simulation, fitting, visualization, and reporting. The implementation uses an analytically integrated exponential--Gaussian response model, configurable source and sample components, constrained optimization, optional least-squares refinement, and machine-readable output of fit results, correlation matrices, and fitted curves. Validation on fully synthetic spectra with known ground-truth parameters shows accurate recovery of the simulated lifetimes, intensities, detector full width at half maximum, prompt shift, and background.
\end{abstract}

%


\maketitle

\section{Introduction}

Positron Annihilation Lifetime Spectroscopy (PALS) \cite{Dupasquier1979,Siegel1980}
is a well-established experimental technique widely used for the characterization
of defects, free-volume distributions, and electronic environments in condensed
matter systems. By analyzing the time interval between positron implantation and
annihilation, PALS provides microscopic information on materials ranging from
metals and semiconductors to polymers and porous materials
\cite{Jean1990,Gidley2006,Tuomisto2013}. The interpretation of PALS measurements relies on
the extraction of characteristic lifetimes and intensities associated with
distinct annihilation channels, enabling the investigation of vacancy-type
defects, open-volume structures, and radiation-induced damage. In particular,
PALS has become an important tool for the characterization of irradiation-induced
defects in tungsten and other fusion-relevant materials
\cite{Taylor2021,Chatzikos2022,Papadakis2023}. Consequently, accurate and
reliable spectrum fitting constitutes a central component of the PALS analysis
procedure.

In practical applications, experimentally measured lifetime spectra are commonly modeled as sums of exponential decay components convoluted with the finite detector resolution function and combined with background contributions. The resulting inverse problem is often nontrivial due to strong parameter correlations, limited counting statistics, detector broadening effects, and the presence of multiple closely spaced lifetime components. Consequently, robust numerical fitting procedures and careful uncertainty estimation are essential for obtaining physically meaningful results. Several software packages have been developed for PALS analysis with the purpose of extracting, from the experimental data, the lifetimes and relative weights of the components forming the PALS spectrum. 

Traditional PALS analysis is commonly performed using discrete multi-exponential fitting methods. Alternative approaches include numerical Laplace inversion, as implemented in CONTIN \cite{Gregory1990,Gregory1991}, and maximum-entropy techniques such as MELT \cite{Shukla1993}, which aim to recover continuous lifetime distributions from experimental spectra. Dedicated software packages for multi-exponential analysis of PALS spectra have been developed, including POSITRONFIT \cite{Kirkegaard1972}, the LT family of programs \cite{Kansy1996,Dryzek1996,Giebel2010,Giebel2012,LTWebsite}, PALSfit and its successor PALSfit3 \cite{Olsen2007,Olsen2019,PALSfitWebsite,PALSfitReport}, and PAScual \cite{PascualIzarra2009,PAScualWebsite}. These programs remain among the most widely used tools for the evaluation of positron lifetime spectra and have played a significant role in establishing standard analysis methodologies within the PALS community.

Despite the success of existing PALS analysis packages, the increasing adoption of open scientific software, reproducible computational workflows, and modern data-analysis ecosystems has created a demand for flexible and extensible tools that can be readily integrated into automated analysis pipelines. Furthermore, the widespread availability of high-level scientific computing environments provides new opportunities for combining advanced optimization algorithms, uncertainty quantification methods, and user-friendly interfaces within a unified framework.

Motivated by these developments, we present fitPALSpectra \cite{fitPALSpectraRepo}, an open-source Python framework for the analysis and fitting of positron annihilation lifetime spectra. The software supports multi-component lifetime models, detector-resolution convolution, source corrections, parameter constraints, uncertainty estimation, and configurable optimization workflows within a modular and extensible architecture. The framework combines deterministic local optimization methods with stochastic global search strategies, such as dual annealing \cite{Xiang2013}, to reduce sensitivity to initial parameter estimates and improve robustness against convergence to suboptimal solutions. Constrained fitting problems may be handled through bound constraints and optional augmented-Lagrangian formulations \cite{Birgin2014}, enabling the incorporation of physically motivated parameter restrictions when required.

Particular emphasis is placed on reproducibility, transparent parameter handling, and comprehensive export of fitting results, including correlation matrices, residual diagnostics, and machine-readable output formats. Unlike many existing packages, fitPALSpectra is distributed as an open-source Python library and command-line application, facilitating integration with modern scientific workflows, automated analysis pipelines, and reproducible computational environments while maintaining compatibility with established PALS fitting methodologies.

The paper is organized as follows. Section 2 introduces the mathematical model used for the description of lifetime spectra and detector resolution effects. Section 3 outlines the software architecture and implementation details of the framework. Validation with synthetic spectra is presented in Section 4. A literature-motivated tungsten-like synthetic benchmark is discussed in Section 5. Finally, conclusions and possible future extensions are summarized in Section 6.

\section{Mathematical model}

The fitted spectrum is described in a discrete-channel form that follows the way multichannel-analyzer data are stored and processed. Following the channel-based convention used in PALSfit \cite{PALSfitReport}, the time origin is arbitrary and only relative shifts are physically meaningful. For a selected fitting window of \(n\) channels, the lower edge of channel \(i\) is defined as
\begin{equation}
t_i = (k_i-k_{\mathrm{peak}})\Delta t,
\end{equation}
where \(k_i\) is the one-based channel number, \(k_{\mathrm{peak}}\) is the prompt-peak channel, and \(\Delta t\) is the channel width. The model count for channel \(i\) is therefore evaluated over the interval \([t_i,t_i+\Delta t]\); choosing a channel-center convention instead would only add a constant half-channel shift that is absorbed by the fitted time-offset parameters. The detector resolution function is represented as a sum of \(k_g\) Gaussian components with normalized IRF weights \(\omega_p\), shifts \(\Delta_p\), and full widths at half maximum \(\mathrm{FWHM}_p\). Each full width at half maximum is converted internally to a Gaussian standard deviation,
\begin{equation}
\sigma_p = \frac{\mathrm{FWHM}_p}{2\sqrt{2\ln 2}}.
\end{equation}

Before channel integration, the integral of the response of one exponential lifetime component convoluted with one Gaussian resolution component must be calculated. The program avoids numerical quadrature by using the analytic primitive of the channel-integrated exponential--Gaussian response. We define $\Psi(u;\tau,\sigma)$ as:
\begin{equation}
\begin{split}
\Psi(u;\tau,\sigma)=\frac{1}{2}\biggl[
&\exp\left(-\frac{u}{\tau}+\frac{\sigma^2}{2\tau^2}\right)
\operatorname{erfc}\left(\frac{\sigma^2-u\tau}{\sqrt{2}\sigma\tau}\right)\\
&+\operatorname{erfc}\left(\frac{u}{\sqrt{2}\sigma}\right)
\biggr].
\end{split}
\label{eq:psi}
\end{equation}
With this definition, the bin-integrated model matrix for the component family \(q\), either source or sample, is:
\begin{equation}
\begin{split}
M^{(q)}_{ij}=\sum_{p=1}^{k_g}\omega_p\{&
\Psi(t_i-T_0-\Delta_p;\tau^{(q)}_j,\sigma_p)\\
&-\Psi(t_i+\Delta t-T_0-\Delta_p;\tau^{(q)}_j,\sigma_p)\}.
\end{split}
\label{eq:matrix}
\end{equation}
Here \(T_0\) is the global time-zero offset. Only the combination \(T_0+\Delta_p\) is identifiable for a single Gaussian IRF; in the reported fits we fix \(T_0=0\) and fit \(\Delta_1\) as the prompt-shift parameter. The shape entering the spectrum is the sum of source and sample contributions,
\begin{equation}
S_i=s_{\mathrm{src}}\sum_j I^{\mathrm{src}}_jM^{\mathrm{src}}_{ij}+s_{\mathrm{samp}}\sum_j I^{\mathrm{samp}}_jM^{\mathrm{samp}}_{ij},
\end{equation}
where \(s_{\mathrm{src}}\) and \(s_{\mathrm{samp}}\) are the source and sample fractions, while \(I^{\mathrm{src}}_j\) and \(I^{\mathrm{samp}}_j\) are the source and sample component intensities. This notation is also used in the active-parameter correlation matrices, where \(\tau^{\mathrm{samp}}_j\), \(\Delta_1\), \(\mathrm{FWHM}_1\), and \(B\) denote the fitted sample lifetimes, prompt shift, single-Gaussian resolution width, and constant background. The final model count in channel \(i\) is
\begin{equation}
f_i = B + NS_i,
\qquad
N = \frac{\sum_i(y_i-B)}{\sum_i S_i},
\label{eq:model}
\end{equation}
with constant background \(B\) and normalization \(N\) determined from the selected fitting window.

The optimized objective function is the weighted chi-square,
\begin{equation}
\chi^2=\sum_{i=1}^{n}\left(\frac{y_i-f_i}{d_i}\right)^2,
\qquad
d_i=\max\left(\sqrt{|y_i|},1\right),
\end{equation}
where \(y_i\) are the measured counts. The reported reduced chi-square is
\begin{equation}
\chi^2_{\mathrm{red}}=\frac{\chi^2}{\nu},
\qquad
\nu=n-p_{\mathrm{act}},
\end{equation}
where \(\nu\) is the number of degrees of freedom and \(p_{\mathrm{act}}\) is the number of parameters with active fit flags. This is a reporting convention; equality constraints and the profiled normalization are not used to redefine the reported degrees of freedom. Equality constraints enforce normalized sample intensities, source intensities, source/sample fractions, and Gaussian IRF weights:
\begin{equation}
\begin{aligned}
\sum_j I^{\mathrm{samp}}_j &= 1,
&\sum_j I^{\mathrm{src}}_j &= 1,\\
s_{\mathrm{src}}+s_{\mathrm{samp}} &= 1,
&\sum_p\omega_p &= 1.
\end{aligned}
\label{eq:constraints}
\end{equation}
As mentioned above, these constraints can be taken into account using augmented-Lagrangian formulations \cite{Birgin2014}. The program can keep any subset of parameters fixed, so the same parameter vector supports unconstrained exploratory fits, constrained production fits, and fits with externally supplied source-correction values.

\section{Software architecture and workflow}

The code is organized as a small Python package, \texttt{pals}, with two command-line entry points for simulation and fitting. Configuration is stored in INI files so that the full analysis can be reproduced from the spectrum file, the configuration file, and the generated output directory. The main modules are:
\begin{itemize}
\item \texttt{config.py}, which reads INI files and parses scalar or list-valued options;
\item \texttt{parameters.py}, which builds the ordered parameter vector, fit flags, labels, and lifetime-sorted component views;
\item \texttt{preprocess.py}, which reads spectrometer-style spectra, optionally extracts header metadata, and selects the fitting window;
\item \texttt{model.py}, which evaluates Eqs.~(\ref{eq:psi})--(\ref{eq:model}) and the normalized residual vector;
\item \texttt{fit.py}, which provides local and global optimization routines, constraint handling, least-squares refinement, and correlation estimates;
\item \texttt{plots.py} and \texttt{reports.py}, which generate static figures, interactive HTML plots, text reports, CSV summaries, and JSON exports.
\end{itemize}

The fitting workflow begins by loading the spectrum and the analysis configuration. If the input file contains a four-line spectrometer header, the channel width and initial resolution estimate can be read directly from the file; otherwise they are taken from the configuration. The fitting window can be selected manually, by a fraction of the prompt-peak height, or by an automatic Gaussian estimate of the prompt position followed by a moving-average tail threshold. The extracted window is then converted to time, count, and uncertainty arrays.

Parameter values and activity flags are specified separately for the source correction, sample components, Gaussian resolution function, mixture fractions, time-zero offset, and background. The fit can be performed with Nelder--Mead, Powell, or dual annealing optimization. In the present implementation, Nelder--Mead is used as a deterministic local optimizer for a fixed set of starting parameters, whereas dual annealing (DA) is a stochastic, bound-constrained global search that does not require a user-selected starting point (a discussion about this can be found in Ref. \cite{Pavlou2025}). When constraints are enabled, an augmented-Lagrangian term penalizes deviations from Eq.~(\ref{eq:constraints}). The resulting parameter set can then be refined with a bounded trust-region-reflective (TRF) least-squares step. Parameter uncertainties and correlation matrices are estimated from the TRF Jacobian using a pseudoinverse of \(J^TJ\). For constrained TRF refinements, this Jacobian is computed from the augmented residual vector, which contains both the data residuals and the scaled equality-constraint residuals. The reported \(\chi^2_{\mathrm{red}}\), however, is computed separately from the data residuals only, using the reporting convention \(\nu=n-p_{\mathrm{act}}\). These uncertainties quantify the local statistical sensitivity of the selected model and weighting scheme; they do not include systematic uncertainty associated with model choice, calibration, or source-correction assumptions.

Every fit creates a timestamped output directory. The directory contains the copied configuration file, a human-readable text report, a row appended to the global CSV summary, the fitted curve and residuals, covariance and correlation matrices in both text and JSON formats, heat maps for active fitted parameters, and static and interactive plots of the spectrum and fit. The companion simulation script uses the same model function to generate spectrometer-style synthetic spectra with optional Poisson noise and writes a \texttt{truth.json} file containing the exact simulated parameters.

\section{Validation with synthetic spectra}

\begin{table}[pos=!htbp]
\caption{Comparison of the known simulation parameters, fitPALSpectra, and LT10 for the first synthetic spectrum. LT10 intensities and intensity deviations are converted from percent to fractional units. Uncertainties are one-standard-deviation estimates where reported and LT10 entries shown without a \(\pm\) value are values for which the LT10 output did not provide a separate uncertainty.}\label{tbl:synthetic}
\scriptsize
\begin{tabular*}{\tblwidth}{@{} LLLL@{}}
\toprule
Quantity & True value & fitPALSpectra & LT10 \\
\midrule
Reduced fit statistic & -- & 0.9755 & 0.9675 \\
Source intensity 1 & 0.9500 & \(0.9515\pm0.0020\) & 0.9492 \\
Source intensity 2 & 0.0500 & \(0.0485\pm0.0020\) & \(0.0508\pm0.0050\) \\
Sample intensity 1 & 0.6000 & \(0.6071\pm0.0092\) & 0.6011 \\
Sample intensity 2 & 0.4000 & \(0.3929\pm0.0092\) & \(0.3989\pm0.0091\) \\
Sample lifetime 1 (ns) & 0.1500 & \(0.1523\pm0.0033\) & \(0.1501\pm0.0013\) \\
Sample lifetime 2 (ns) & 0.5000 & \(0.5058\pm0.0086\) & \(0.4989\pm0.0033\) \\
Prompt shift (ns) & -0.0100 & \(-0.0103\pm0.0007\) & \(-0.00989\pm0.00078\) \\
IRF FWHM (ns) & 0.2700 & \(0.2694\pm0.0013\) & \(0.27032\pm0.00006\) \\
Background & 100.0 & \(99.37\pm0.57\) & \(99.67\pm0.55\) \\
\bottomrule
\end{tabular*}
\end{table}

A synthetic test spectrum was generated from the simulation configuration file. The simulated spectrum used 6000 channels, \(\Delta t=0.0065\) ns, two million signal counts, a background of 100 counts/channel, and a prompt peak at channel 3000. The detector resolution was a single Gaussian with \(\mathrm{FWHM}_1=0.270\) ns and \(\Delta_1=-0.010\) ns. The source and sample fractions were both 0.5. The source correction contained fixed lifetimes of 0.5 ns and 2.5 ns with intensities 0.95 and 0.05, while the sample contained two components with lifetimes 0.15 ns and 0.50 ns and intensities 0.60 and 0.40.

The synthetic spectrum was then fitted with DA followed by a TRF refinement. The fit used the header channel width, a manual fitting window from channels 2980 to 4500, constrained intensity normalization, bounded TRF
refinement, and a two-component source/sample model. The program converged, with \(\chi^2_{\mathrm{red}}=0.9755\) (see Table \ref{tbl:synthetic}). A Nelder-Mead run followed by the
same TRF refinement produced practically the same results.

\begin{figure*}[pos=!tbp]
\centering
\begin{minipage}[t]{.48\textwidth}
\centering
\includegraphics[width=\linewidth]{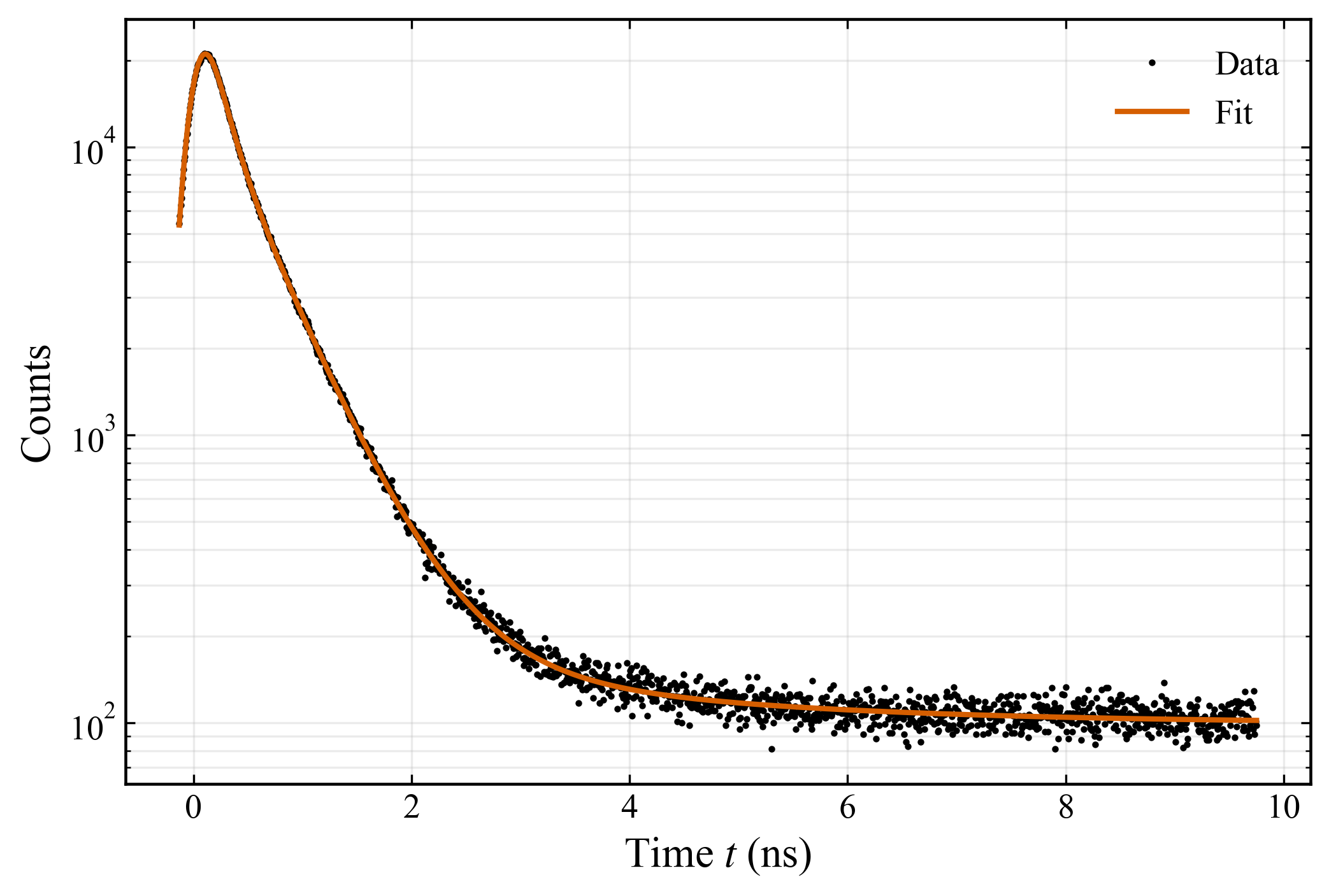}
\end{minipage}\hfill
\begin{minipage}[t]{.48\textwidth}
\centering
\includegraphics[width=.82\linewidth]{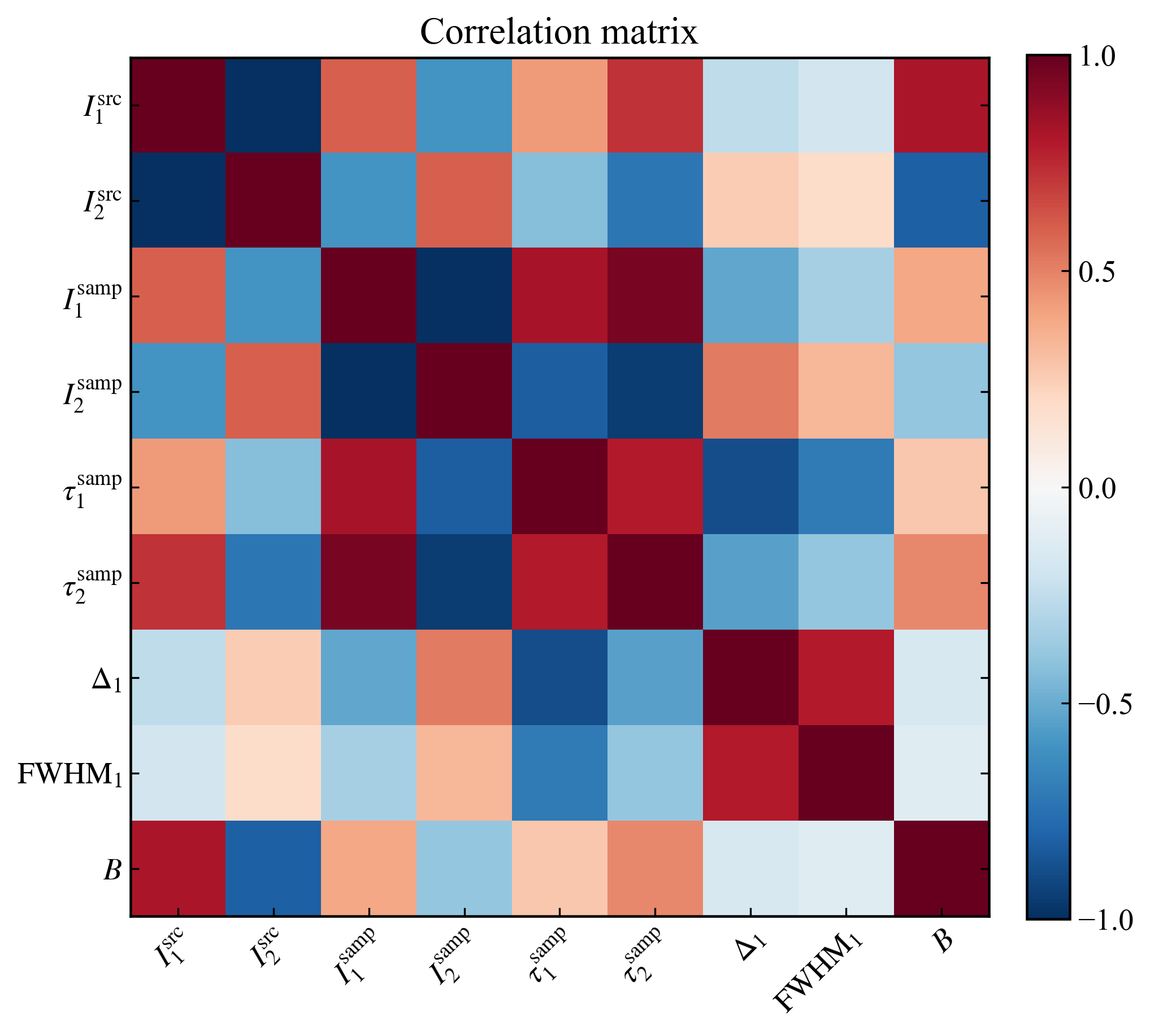}
\end{minipage}
\caption{Generic synthetic benchmark: fitted PALS spectrum (left) and active-parameter correlation matrix (right). The correlation labels follow the notation of Eqs.~(\ref{eq:matrix})--(\ref{eq:constraints}); only parameters allowed to vary in the fit are shown.}
\label{fig:synthetic-results}
\end{figure*}

The same spectrum was also fitted independently with LT10. The LT10 spreadsheet reports intensities in percent; all LT10 intensities and intensity deviations in Table~\ref{tbl:synthetic} have therefore been converted to fractional units. LT10 used the same zero, start, and stop channels (3000, 2980, and 4500, respectively) and reported a fit statistic of 0.9675. The small differences between LT10 and fitPALSpectra are comparable to the reported local uncertainties and are expected for independent implementations with different internal normalization and error-reporting conventions. Table~\ref{tbl:synthetic} compares the simulation truth, the DA+TRF fitPALSpectra result, and the LT10 result, while Fig.~\ref{fig:synthetic-results} shows the fitted curve and active-parameter correlation matrix.

\section{Literature-motivated tungsten-like synthetic case}

\begin{table}[pos=!htbp]
\caption{Comparison of the known simulation parameters, fitPALSpectra, and LT10 for the literature-motivated tungsten-like synthetic benchmark. LT10 intensities and intensity deviations are converted from percent to fractional units. Uncertainties are one-standard-deviation estimates where reported; LT10 entries shown without a \(\pm\) value are values for which the LT10 output did not provide a separate uncertainty.}\label{tbl:tungsten-synthetic}
\scriptsize
\resizebox{\columnwidth}{!}{%
\begin{tabular}{@{}llll@{}}
\toprule
Quantity & True value & fitPALSpectra & LT10 \\
\midrule
Reduced fit statistic & -- & 0.9788 & 0.9750 \\
Source intensity 1 & 0.9500 & \(0.9510\pm0.0014\) & 0.9535 \\
Source intensity 2 & 0.0500 & \(0.0490\pm0.0014\) & \(0.0465\pm0.0042\) \\
Sample intensity 1 & 0.7000 & \(0.7041\pm0.0128\) & 0.7161 \\
Sample intensity 2 & 0.3000 & \(0.2959\pm0.0128\) & \(0.2839\pm0.0175\) \\
Sample lifetime 1 (ns) & 0.1100 & \(0.1117\pm0.0033\) & \(0.1153\pm0.0027\) \\
Sample lifetime 2 (ns) & 0.3500 & \(0.3521\pm0.0097\) & \(0.3632\pm0.0077\) \\
Prompt shift (ns) & -0.0100 & \(-0.0104\pm0.0010\) & \(-0.0113\pm0.0024\) \\
IRF FWHM (ns) & 0.2700 & \(0.2693\pm0.0014\) & \(0.2673\pm0.0023\) \\
Background & 100.0 & \(99.38\pm0.51\) & \(101.12\pm0.33\) \\
\bottomrule
\end{tabular}%
}
\end{table}

To illustrate a physically motivated application without using unpublished or laboratory-specific measurements, a second fully synthetic spectrum was constructed using lifetime values representative of tungsten studies. Reported values for defect-free tungsten are typically close to \(100\)--\(110\) ps, while vacancy-type defects, vacancy clusters, and vacancy--impurity complexes are associated with longer lifetimes whose values depend on defect size, composition, and charge state \cite{Chatzikos2022,Yabuuchi2023}. These literature ranges were used only to define a reproducible benchmark spectrum, not to assign a unique defect identity to the synthetic components.

\begin{figure*}[pos=!tbp]
\centering
\begin{minipage}[t]{.48\textwidth}
\centering
\includegraphics[width=\linewidth]{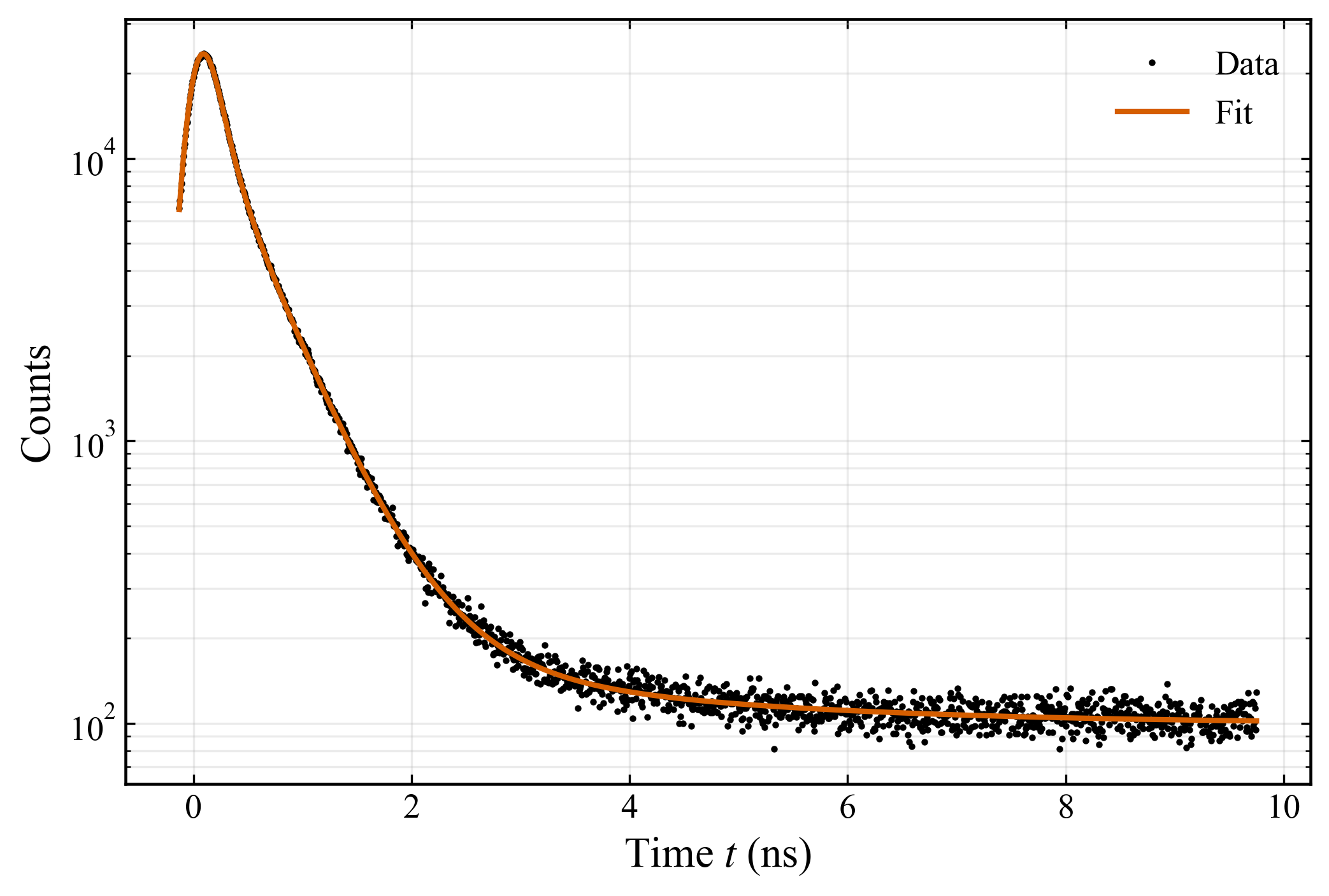}
\end{minipage}\hfill
\begin{minipage}[t]{.48\textwidth}
\centering
\includegraphics[width=.82\linewidth]{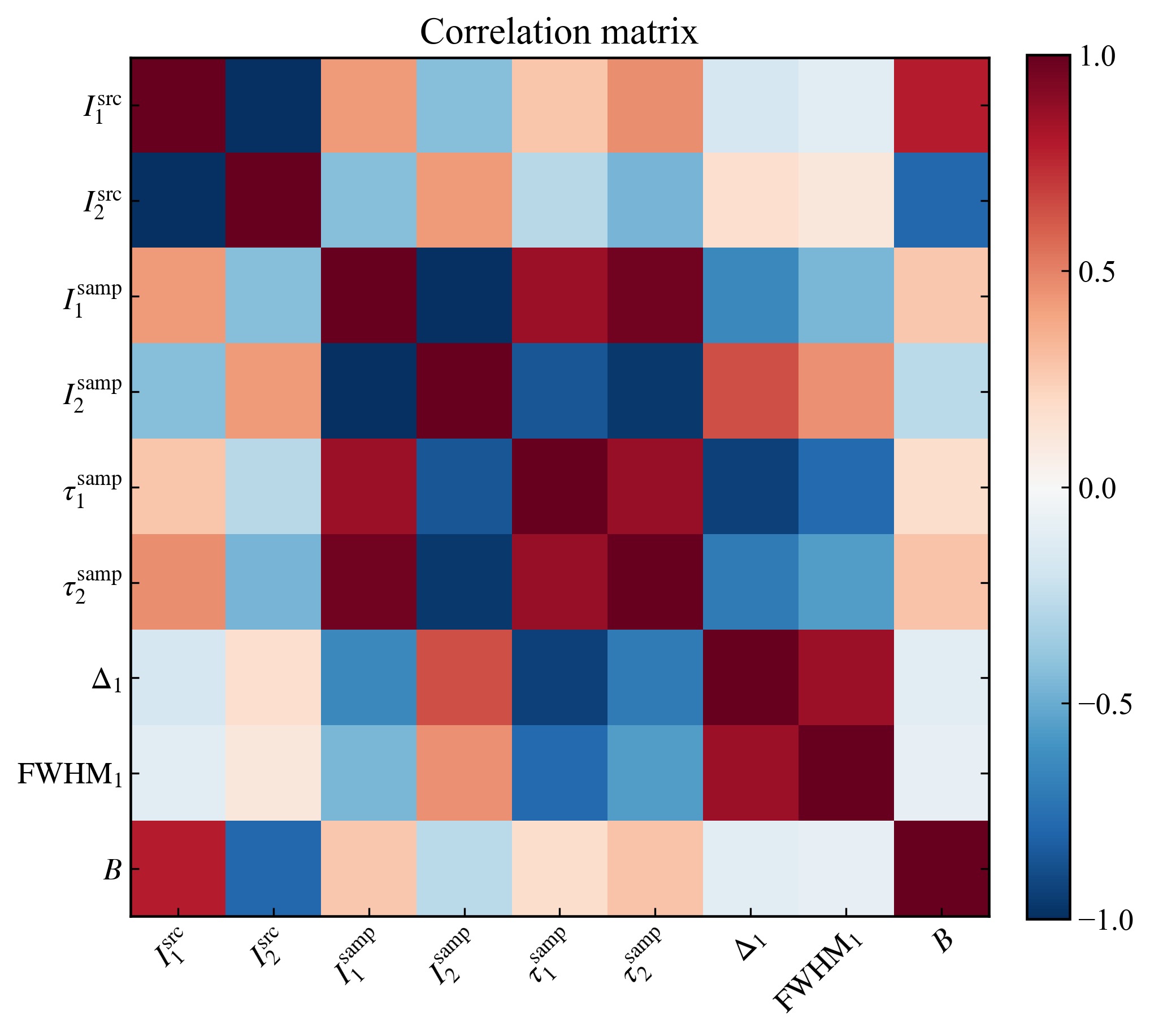}
\end{minipage}
\caption{Literature-motivated tungsten-like synthetic benchmark: fitted PALS spectrum (left) and active-parameter correlation matrix (right), using the same notation as Fig.~\ref{fig:synthetic-results}. The strong anticorrelation between complementary intensities and the correlations among \(\tau^{\mathrm{samp}}_j\), prompt shift \(\Delta_1\), resolution width \(\mathrm{FWHM}_1\), and background \(B\) illustrate why direct comparisons with established tools should use identical fitting windows, source corrections, constraints, and parameter bounds.}
\label{fig:tungsten-like-results}
\end{figure*}

The simulated tungsten-like case used two sample components: a bulk-like component with lifetime \(0.110\) ns and intensity 0.70, and a longer defect-like component with lifetime \(0.350\) ns and intensity 0.30. The detector response, channel width, background, source correction, and fitting window were chosen to match the same numerical workflow used for the generic synthetic validation. This case was fitted with DA+TRF resulting in convergence with \(\chi^2_{\mathrm{red}}=0.9788\). The same spectrum was also fitted independently with LT10, using the same zero, start, and stop channels, and LT10 reported a fit statistic of 0.9750. Table~\ref{tbl:tungsten-synthetic} compares the simulation truth, the fitPALSpectra result, and the LT10 result, with LT10 intensities converted from percent to fractional units as in Table~\ref{tbl:synthetic}. Fig.~\ref{fig:tungsten-like-results} shows the fitted spectrum and active-parameter correlation matrix for the same benchmark. The optimizer agreement is reassuring but not, by itself, a strong benchmark result; the more useful comparison is that the independent LT10 fit recovers similar sample lifetimes and intensities for the same synthetic case.

\FloatBarrier
\section{Conclusions and outlook}

fitPALSpectra provides a reproducible, configuration-driven workflow for PALS spectrum simulation and analysis. The current implementation combines an analytic convolution model, source and sample component handling, constrained optimization, bounded least-squares refinement, uncertainty estimation, and comprehensive reporting in text, CSV, JSON, PNG, PDF, and HTML formats. The synthetic validations demonstrate that the workflow can recover known model parameters accurately, while the tungsten-like benchmark shows how literature-motivated cases can be constructed without relying on unpublished or laboratory-specific data.

Future work will focus on broader benchmark comparisons with established PALS analysis programs and the inclusion of the trapping model \cite{Tuomisto2013}. Because the code is modular, these extensions can be added without changing the basic spectrum model, reporting format, or configuration-driven analysis design.

\section*{Data and code availability}

The fitPALSpectra source code, input configuration files, synthetic spectra, generated fit outputs, and plotting data needed to reproduce the fitPALSpectra results reported in this work are available in the project repository at \url{https://github.com/gepavlou/fitPALSpectra}. The repository is distributed under the MIT license and includes a \path{README.md} file with installation instructions, the fixed-name configuration workflow, and instructions for rerunning archived benchmark fits. The exact DA+TRF outputs used for Tables~\ref{tbl:synthetic} and~\ref{tbl:tungsten-synthetic} are stored in \path{outputs/simulated_spectrum_1_20260606_221015} and \path{outputs/simulated_spectrum_W_20260607_024532}, respectively; the corresponding \path{pals_used.ini} files in those directories are the authoritative fitting configurations for the reported runs.
\FloatBarrier
\bibliographystyle{cas-model2-names}


\begin{thebibliography}{10}
\bibitem{Dupasquier1979}
A.~Dupasquier and P.~Hautoj\"arvi (Eds.),
\emph{Positrons in Solids},
Springer-Verlag, Berlin, 1979.

\bibitem{Siegel1980}
R.~W. Siegel,
``Positron Annihilation Spectroscopy,''
\emph{Annual Review of Materials Science}
\textbf{10}(1), 393--425 (1980).
\href{https://doi.org/10.1146/annurev.ms.10.080180.002141}
{doi:10.1146/annurev.ms.10.080180.002141}.


\bibitem{Jean1990}
Y.~C. Jean,
``Positron annihilation spectroscopy for chemical analysis: A novel probe for microstructural analysis of polymers,''
\emph{Microchemical Journal}
\textbf{42}, 72--102 (1990).
\href{https://doi.org/10.1016/0026-265x(90)90027-3}{doi:10.1016/0026-265x(90)90027-3}

\bibitem{Gidley2006}
D.~W. Gidley, H.-G. Peng, and R.~S. Vallery,
``Positron annihilation as a method to characterize porous materials,''
\emph{Annual Review of Materials Research}
\textbf{36}, 49--79 (2006).
\href{https://doi.org/10.1146/annurev.matsci.36.111904.135144}{doi:10.1146/annurev.matsci.36.111904.135144}.

\bibitem{Tuomisto2013}
F.~Tuomisto and I.~Makkonen,
``Defect identification in semiconductors with positron annihilation: Experiment and theory,''
\emph{Reviews of Modern Physics}
\textbf{85}(4), 1583--1631 (2013).
\href{https://doi.org/10.1103/RevModPhys.85.1583}
{doi:10.1103/RevModPhys.85.1583}.

\bibitem{Taylor2021}
C.~N. Taylor, M.~Shimada, J.~M. Watkins,
X.~Hu, and Y.~Oya,
``Neutron irradiated tungsten bulk defect characterization by positron annihilation spectroscopy,''
\emph{Nuclear Materials and Energy}
\textbf{26}, 100936 (2021).
\href{https://doi.org/10.1016/j.nme.2021.100936}
{doi:10.1016/j.nme.2021.100936}.

\bibitem{Chatzikos2022}
V.~Chatzikos, K.~Mergia, G.~Bonny,
D.~Terentyev, D.~Papadakis,
G.~E. Pavlou, and S.~Messoloras,
``Positron annihilation spectroscopy investigation of defects in neutron irradiated tungsten materials,''
\emph{International Journal of Refractory Metals and Hard Materials}
\textbf{105}, 105838 (2022).
\href{https://doi.org/10.1016/j.ijrmhm.2022.105838}
{doi:10.1016/j.ijrmhm.2022.105838}.

\bibitem{Papadakis2023}
D.~Papadakis, K.~Mergia, E.~Manios, V.~Chatzikos,
S.~Dellis, and S.~Messoloras,
``Post neutron irradiation annealing and defect evolution in single crystal tungsten,''
\emph{Nuclear Materials and Energy}
\textbf{34}, 101357 (2023).
\href{https://doi.org/10.1016/j.nme.2022.101357}
{doi:10.1016/j.nme.2022.101357}.

\bibitem{Gregory1990}
R.~B. Gregory and Y.~Zhu,
``Analysis of positron annihilation lifetime data by numerical Laplace inversion with the program CONTIN,''
\emph{Nuclear Instruments and Methods in Physics Research Section A}
\textbf{290}(1), 172--182 (1990).
\href{https://doi.org/10.1016/0168-9002(90)90358-D}
{doi:10.1016/0168-9002(90)90358-D}.

\bibitem{Gregory1991}
R.~B. Gregory,
``Free-volume and pore size distributions determined by numerical Laplace inversion of positron annihilation lifetime data,''
\emph{Journal of Applied Physics}
\textbf{70}(9), 4665--4670 (1991).
\href{https://doi.org/10.1063/1.349057}
{doi:10.1063/1.349057}.

\bibitem{Shukla1993}
A.~Shukla, M.~Peter, and L.~Hoffmann,
``Analysis of positron lifetime spectra using quantified maximum entropy and a general linear filter,''
\emph{Nuclear Instruments and Methods in Physics Research Section A}
\textbf{335}(1--2), 310--317 (1993).
\href{https://doi.org/10.1016/0168-9002(93)90286-Q}
{doi:10.1016/0168-9002(93)90286-Q}.

\bibitem{Kirkegaard1972}
P.~Kirkegaard and M.~Eldrup,
``POSITRONFIT: A versatile program for analysing positron lifetime spectra,''
\emph{Computer Physics Communications}
\textbf{3}(3), 240--255 (1972).
\href{https://doi.org/10.1016/0010-4655(72)90070-7}
{doi:10.1016/0010-4655(72)90070-7}.


\bibitem{Kansy1996}
J.~Kansy,
``Microcomputer program for analysis of positron annihilation lifetime spectra,''
\emph{Nuclear Instruments and Methods in Physics Research Section A}
\textbf{374}(2), 235--244 (1996).
\href{https://doi.org/10.1016/0168-9002(96)00075-7}
{doi:10.1016/0168-9002(96)00075-7}.

\bibitem{Dryzek1996}
J.~Dryzek and J.~Kansy,
``Comparison of three programs: Positronfit, Resolution and LT used for deconvolution of positron lifetime spectra,''
\emph{Nuclear Instruments and Methods in Physics Research Section A}
\textbf{380}(3), 576--581 (1996).
\href{https://doi.org/10.1016/0168-9002(96)00586-4}
{doi:10.1016/0168-9002(96)00586-4}.

\bibitem{Giebel2010}
D.~Giebel and J.~Kansy,
``A New Version of LT Program for Positron Lifetime Spectra Analysis,''
\emph{Materials Science Forum}
\textbf{666}, 138--141 (2010).
\href{https://doi.org/10.4028/www.scientific.net/MSF.666.138}{doi:10.4028/www.scientific.net/msf.666.138}.

\bibitem{Giebel2012}
D.~Giebel and J.~Kansy,
``LT10 Program for Solving Basic Problems Connected with Defect Detection,''
\emph{Physics Procedia}
\textbf{35}, 122--127 (2012).
\href{https://doi.org/10.1016/j.phpro.2012.06.022}
{doi:10.1016/j.phpro.2012.06.022}.

\bibitem{LTWebsite}
\href{https://ifj.edu.pl/private/jdryzek/page_r29.html}{LT10 software homepage},
accessed June 2026.

\bibitem{Olsen2007}
J.~V. Olsen, P.~Kirkegaard, N.~J. Pedersen, and M.~Eldrup,
``PALSfit: A new program for the evaluation of positron lifetime spectra,''
\emph{physica status solidi (c)}
\textbf{4}(10), 4004--4006 (2007).
\href{https://doi.org/10.1002/pssc.200675868}{doi:10.1002/pssc.200675868}.

\bibitem{Olsen2019}
J.~V. Olsen, P.~Kirkegaard, and M.~Eldrup,
``Analysis of positron lifetime spectra using the PALSfit3 program,''
in \emph{International Conference on Science and Applied Science (ICSAS) 2019},
AIP Conference Proceedings \textbf{2202}, 040005 (2019).
\href{https://doi.org/10.1063/1.5135837}{doi:10.1063/1.5135837}.

\bibitem{PALSfitWebsite}
\href{https://palsfit.dk/index.htm}{PALSfit3 software homepage},
accessed June 2026.

\bibitem{PALSfitReport}
\href{https://backend.orbit.dtu.dk/ws/files/307134294/PALSfit3_final_2022.pdf}{Official PALSfit3 report},
accessed June 2026.

\bibitem{PascualIzarra2009}
C.~Pascual-Izarra, A.~W. Dong, S.~J. Pas,
A.~J. Hill, B.~J. Boyd, and C.~J. Drummond,
``Advanced fitting algorithms for analysing positron annihilation lifetime spectra,''
\emph{Nuclear Instruments and Methods in Physics Research Section A}
\textbf{603}(3), 456--466 (2009).
\href{https://doi.org/10.1016/j.nima.2009.01.205}{doi:10.1016/j.nima.2009.01.205}.

\bibitem{PAScualWebsite}
\href{https://sourceforge.net/projects/pascual/}{PAScual software project},
accessed June 2026.

\bibitem{fitPALSpectraRepo}
G.~E. Pavlou,
fitPALSpectra software repository,
\url{https://github.com/gepavlou/fitPALSpectra},
accessed June 2026.

\bibitem{Xiang2013}
Y.~Xiang, S.~Gubian, B.~Suomela, and J.~Hoeng,
``Generalized Simulated Annealing for Global Optimization: The GenSA Package,''
\emph{The R Journal}
\textbf{5}(1), 13--28 (2013).
\href{https://doi.org/10.32614/RJ-2013-002}
{doi:10.32614/RJ-2013-002}.

\bibitem{Birgin2014}
E.~G. Birgin and J.~M. Mart\'{\i}nez,
\emph{Practical Augmented Lagrangian Methods for Constrained Optimization},
Society for Industrial and Applied Mathematics (SIAM), Philadelphia, PA, 2014.
\href{https://doi.org/10.1137/1.9781611973365}
{doi:10.1137/1.9781611973365}.

\bibitem{Pavlou2025}
G. E. Pavlou, V. Pavlidou, and V. Harmandaris,
Reconstructing the Magnetic Field in an Arbitrary Domain via Data-Driven Bayesian Methods and Numerical Simulations,
\textit{Computation} \textbf{13}, 37 (2025).
\href{https://doi.org/10.3390/computation13020037}{10.3390/computation13020037}.

\bibitem{Yabuuchi2023}
A.~Yabuuchi,
``Inverse change in positron lifetimes of vacancies in tungsten by binding of interstitial impurity atoms to a vacancy: A first-principles study,''
\textit{Nucl. Mater. Energy} \textbf{34}, 101364 (2023).
\href{https://doi.org/10.1016/j.nme.2023.101364}{doi:10.1016/j.nme.2023.101364}

\end{thebibliography}


%
%

\end{document}